\pdfoutput=1  
\documentclass{article}

\PassOptionsToPackage{numbers, compress}{natbib}
\usepackage[final, dblblindworkshop]{neurips_2026}

\workshoptitle{AI for Science Workshop (NeurIPS 2026)}

\makeatletter
\renewcommand{\@noticestring}{Accepted at the AI for Science workshop (NeurIPS 2026).}
\makeatother

\usepackage[utf8]{inputenc}
\usepackage[T1]{fontenc}
\usepackage{hyperref}
\hypersetup{
  hidelinks,
  colorlinks=false,
  pdfborder={0 0 0},
  citebordercolor={},
  linkbordercolor={},
  urlbordercolor={},
  filebordercolor={},
  runbordercolor={},
  menubordercolor={}
}
\usepackage{url}
\usepackage{booktabs}
\usepackage{amsfonts}
\usepackage{nicefrac}
\usepackage{microtype}
\usepackage{graphicx}
\usepackage{multirow}

\newcommand{\figdir}{figures}
\newcommand{\FIGEXT}{_tiers.pdf}
\newcommand{\WALLEXT}{_wall.png}

\providecommand{\note}[1]{\GenericError{}{DRAFTING NOTE REACHED THE DOCUMENT: convert it to a comment line before submitting}{}{}}
\providecommand{\budget}[1]{\GenericError{}{DRAFTING BUDGET REACHED THE DOCUMENT: remove it before submitting}{}{}}

\title{The Verification Wall in Microbial Metabolomics}

\author{%
  Takahiko Koyama \\
  Keio University \\
  \texttt{takahiko.koyama@keio.jp} \\
}

\begin{document}

\maketitle

\begin{abstract}
Computational methods now propose structures for gut microbial metabolites faster than any
laboratory can confirm them. We argue that verification, not generation, is the binding
constraint, and that better models will not relieve it. Confirming a structure to the field's
own standard means having the molecule in hand. Automation can lower what that costs per
compound, but the cost stays proportional to the number of structures and never amortizes
across them. Two substitutes are on offer and neither removes that requirement.
In-silico spectral prediction ranks candidates instead of confirming one, and the strongest
current model puts the right structure first 33.5\% of the time on a scaffold-split
benchmark. Any assignment resting on spectral similarity is, by the field's own definition,
made without a reference standard. Microcrystal diffraction does return a structure, but it
needs the compound to crystallize out of a trace mixture, which is a different problem rather
than an easier one. We sorted 10{,}645 computationally inferred metabolite structures by what
verifying each would take. Between 5\% and 37\% have no verification route identified under our
analysis. That spread reflects where the
thresholds are set rather than a measured count of unverifiable compounds, and the
inaccessible ones cluster among the most speculative entries.
Those figures understate the problem, because the gut makes far more compounds than any
database records. HMDB holds only what someone has already thought worth cataloguing, and
structure generators are bounded by nothing at all. The claim is easy to disprove. Show a
route that confirms structures as fast as they are proposed. Until someone does, two problems
stay unsolved. Nobody knows how to spend a verification budget that will always be too small,
or what a biologist may conclude from a structure that was never confirmed.
\end{abstract}

\section{Introduction}
\label{sec:intro}

A typical human gut carries at least 160 common bacterial species, drawn from a pool of
about a thousand \citep{qin2010metahit}. Which species are present differs substantially between
individuals, and most of that difference is not explained by any host trait we measure
\citep{hmp2012structure}. Diet shifts the community too, quickly and repeatably
\citep{david2014diet}. So the enzymes present and the food they act on both vary between
people. The molecules a given gut could make are therefore a combinatorial set, far larger
than what any one sample holds.

\begin{figure}[t]
  \centering
  \IfFileExists{\figdir/fig1\WALLEXT}{%
    \includegraphics[width=\linewidth]{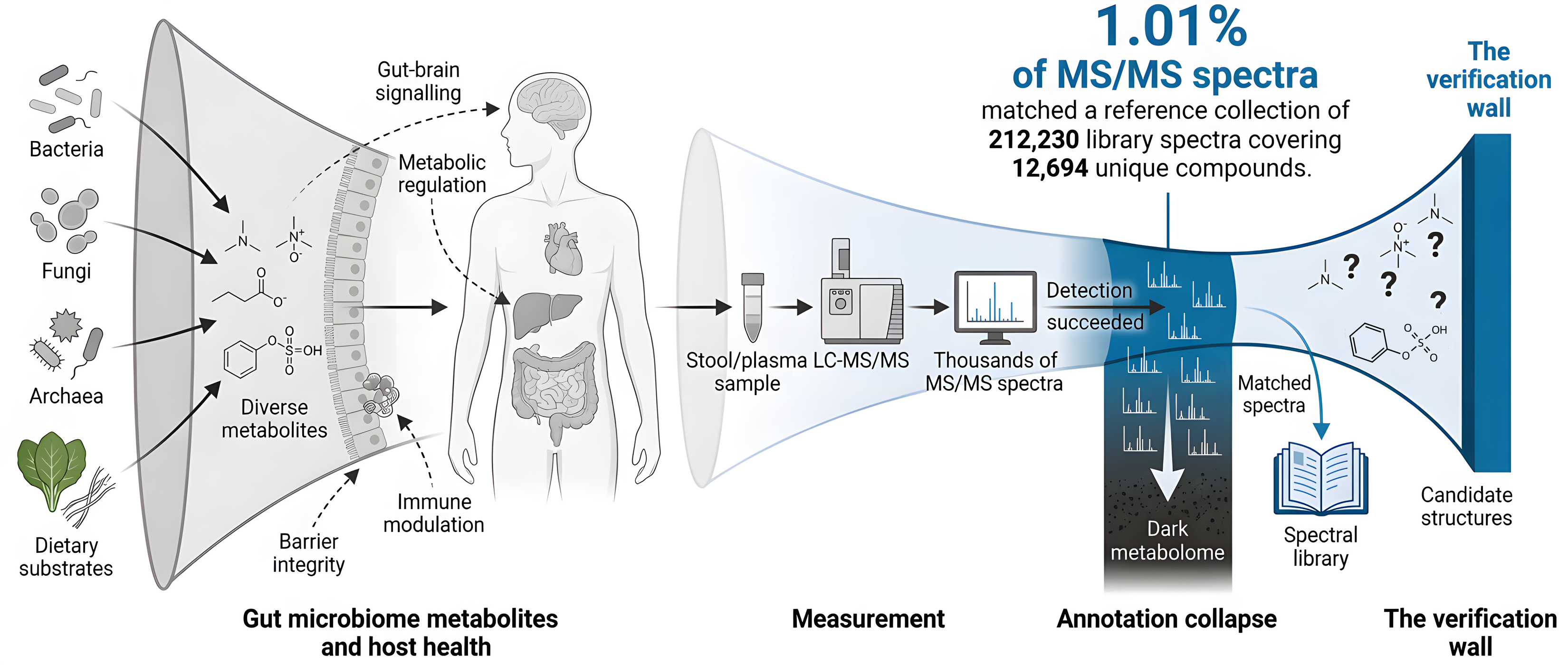}%
  }{%
    \fbox{\parbox[c][1.9in][c]{0.96\linewidth}{\centering\textit{\small Figure~1 placeholder, drop \texttt{\figdir/fig1\_wall.pdf} into place and this box is replaced automatically.}}}%
  }
  \caption{\textbf{The verification wall.} Gut bacteria convert dietary and host compounds
  into small molecules that cross the intestinal epithelium and act on host physiology
  \citep{krautkramer2021multikingdom}. Untargeted LC-MS/MS detects thousands of these
  features per sample, but only 1.01\% of MS/MS spectra in public GNPS datasets match a
  reference collection of 212{,}230 library spectra covering 12{,}694 unique compounds
  \citep{wang2016gnps}. Candidate structures for the remainder are cheap to propose;
  confirming one to MSI level~1 requires an authentic reference standard. Detection and
  generation are not the constraint. Verification is.}\label{fig:wall}
\end{figure}

Detection is not the problem. Untargeted mass spectrometry finds thousands of features in a
single sample. Naming them is what fails. In one study of germ-free and colonized mice,
searching spectral libraries annotated only 8.9\% of the clusters in the molecular network,
where each cluster stands for one putative compound \citep{quinn2020global}. Those
annotations were level 2 or 3 on the Metabolomics Standards Initiative scale, so even the
annotated minority are candidates, not confirmed structures. The same pattern holds across
public repositories, where 1.01\% of MS/MS spectra matched a reference collection of
212{,}230 library spectra covering 12{,}694 unique compounds \citep{wang2016gnps}. The rest is the
dark matter of metabolomics \citep{dasilva2015darkmatter}. Figure~\ref{fig:wall} shows where
it accumulates.

Most of these molecules are probably harmless. Some are not. Gut bacteria convert dietary
phosphatidylcholine into trimethylamine, the host oxidizes it to trimethylamine N-oxide, and
that product promotes cardiovascular disease \citep{wang2011tmao}. We know this for only a
handful of metabolites, each worked out on its own \citep{krautkramer2021multikingdom}.
Trimethylamine N-oxide became useful because someone pinned down its structure, its
microbial source, and its effect on the host. That kind of work does not scale. The cost of
leaving the metabolome unassigned is not an incomplete catalogue. It is that we cannot tell
which unknown matters next.

We argue that the binding constraint here is verification, not generation. Candidates are
cheap. Rule-based enumeration and machine-learning generators produce them in bulk, but
confirming one to level 1 requires an authentic reference standard, meaning a physical
sample of the molecule. We claim that this is out of reach at the scale candidates are generated, and that
the routes offered to close the gap do not close it. Buying or making standards costs money
per compound, so the bill grows with the number of candidates. Microcrystal diffraction
trades that cost for a requirement to crystallize. Spectrum prediction is often called an
alternative, but it ranks candidate structures instead of confirming one, so it is not a
substitute for a physical sample at all. The claim can be falsified. Show any route that
supplies confirmed structures as fast as candidates appear. We do not propose a fix. We
measure the residual and identify two quantities that decide whether it shrinks, namely
retrosynthetic reachability and crystallizability.

\section{What counts as an identified metabolite}
\label{sec:background}

The measurement does not return a structure. Chromatography separates the sample. Each
feature that comes off the column gives two things. One is an accurate mass for the whole
ion. The other is a fragmentation spectrum, giving the masses and intensities of the pieces the
ion breaks into. That
narrows down the structure without fixing it. Isomers have the same exact mass, and knowing
which pieces a molecule breaks into is not the same as knowing how its atoms are joined. In
machine-learning terms, this is a labelling problem where the labels are chemical structures,
the observation only partly determines the label, and no amount of further analysis of the
data recovers the ground truth.

Getting that ground truth is governed by a four-level reporting standard
\citep{sumner2007msi}. Level 1, an identified compound, needs at least two independent and
orthogonal measurements compared against an authentic compound run under identical
conditions. Level 2 is defined as an assignment made \emph{without} a chemical reference
standard, resting on similarity to a library spectrum instead. Level 3 gives only a compound
class, and level 4 means unknown. Later work refines the scheme for high-resolution
instruments \citep{schymanski2014confidence}. But the line this paper turns on is in the
original. Level 1 is the only level that requires having the molecule in hand. Everything
below it is an argument from similarity.

Candidates, by contrast, are cheap. Rule-based enumeration over known biotransformations
\citep{wishart2022biotransformer3} and iterative reaction-network expansion
\citep{jeffryes2015mines} produce plausible structures in bulk. Reference databases already
show the imbalance. Of the 217{,}920 entries in the Human Metabolome Database, 193{,}611,
or 88.8\%, are flagged predicted or expected rather than experimentally detected
\citep{wishart2022hmdb5}. That number counts catalogued entries, not observed spectra, so it
measures the stock of proposals rather than the outcome of any one experiment.

\section{Why verification does not scale}
\label{sec:groundtruth}

Level 1 identification requires measuring an authentic sample alongside the unknown
\citep{sumner2007msi}. There are only two ways to get one. The compound is either purchased or
synthesized.

Buying works more often than we expected. Of the 10{,}645 unverified structures we analyze
in Section~\ref{sec:measurable}, 7{,}241 have a commercial vendor and 3{,}491 have a
supplier that plausibly sells analytical-grade material. But availability tracks how much
attention a compound has already received, not how simple it is. Split the same set by how speculative each entry is, using HMDB's own status field
rather than a judgement of ours. Entries it marks expected are supported by biochemical
reasoning; entries it marks predicted rest on enumeration alone. The better-established compounds have a vendor 68.7\% of the
time, against 29.1\% for the more speculative ones. For analytical-grade suppliers the gap
is wider still, 33.3\% against 2.7\%. The speculative class is small ($n=182$), so we read
this as a direction rather than a precise rate. Some of that gap is circular, since
databases tend to record compounds that are already available. That works against us. It
means 68.7\% overstates how much of a genuinely novel candidate set could be bought.
So the claim we make is the modest one. What is purchasable is what has already been
studied.

When no vendor lists it, the compound must be synthesized, and for metabolite standards
that is routine
rather than exceptional \citep{shanuwilson2024metabolite}. Costs quoted from drug discovery
run to months and tens of thousands of dollars per compound \citep{seal2025aiagents}. Those
are worst cases, for molecules larger than most gut metabolites. Automation raises throughput
without removing the per-compound requirement. Iterative platforms are still benchmarked per bond formed
\citep{wang2024iterative}, and on-demand libraries only contain what their building blocks
allow \citep{tyrin2026onepotcore}. Often the route is simply unknown. An automated search
found none for 5{,}928 of our compounds, many of which are nonetheless purchasable and so
need no route. That figure says as much about the planner's coverage as about the chemistry
(Section~\ref{sec:measurable}).

None of this makes verification impossible. It makes it linear. Each standard confirms
exactly one structure, and confirming it says nothing about the next candidate. So
cost scales with the number of structures to be checked, while generating them costs
almost nothing. There is no amortization here, no point at which paying once makes the rest
cheap.

One reply is that the candidate count is inflated. Enumeration returns what is chemically
possible, and biology permits far less, so ranking candidates by metabolic plausibility
should shrink the set worth checking. Reaction-network context already does this for
annotation \citep{shen2019metdna}, and pathway retrosynthesis does it over enzymatic rules
\citep{delepine2018retropath}. That is the right instinct, but it does not rescue the
arithmetic. These methods rank using curated enzymology, and curation is incomplete in the
same direction as the metabolite catalogue. In the unified human gut protein catalogue,
40\% of proteins carry no functional annotation and more than 70\% of species have no
cultured representative \citep{almeida2021uhgg}. A prior built from known enzymes ranks
known chemistry highly. It cannot show that what it demotes is not there.

That asymmetry is why the field has turned to substitutes for the physical sample, and
why what those substitutes actually deliver is the question that matters.

\section{Prediction ranks rather than confirms}
\label{sec:routes}

Two developments are offered in place of a physical standard. They are not the same kind of
thing. One is not a verification route at all, and saying so precisely matters more than
treating both as partial solutions.

\subsection{Predicted spectra}
\label{sec:insilico}

Spectral prediction has improved quickly. Current models generate fragmentation graphs or
subformula sets rather than binned intensity vectors \citep{goldman2024iceberg,
goldman2023scarf}. Transformers transfer from pretrained molecular representations
\citep{young2024massformer}. The strongest current system conditions on collision energy and
ionization polarity \citep{wang2025iceberg2}. But note what the resulting pipeline
delivers. It takes a set of candidate structures, predicts a spectrum for each, and ranks
them by similarity to the observed spectrum. The output is an ordering over candidates, not
a determination.

The reported accuracies make the consequence concrete. On a random split of the NIST'20 benchmark that
system achieves 40.0\% top-1 retrieval accuracy for [M+H]$^+$ ions (95\% CI 39.2--40.8),
improved from 25.1\% for the previous generation \citep{wang2025iceberg2}. Under a scaffold split, which withholds
structural families to test generalization to chemistry unlike the training set, top-1
accuracy is 33.5\% (95\% CI 33.0--34.0). The scaffold split is the relevant regime for
newly discovered metabolites, so roughly two of every three top-ranked assignments are
wrong there, and the ranking gives no signal distinguishing the third that is right.
Retrieval also presupposes that the true structure appears among the enumerated candidates,
which for a genuinely novel metabolite is not guaranteed.

The deeper problem is categorical rather than quantitative. The field's reporting standard
places any assignment resting on spectral similarity at level~2, explicitly defined as
identification made without a chemical reference standard \citep{sumner2007msi}. A better
predictor produces a better level~2 assignment. It does not produce a level~1 one, because
what level~1 requires is the physical sample the prediction was meant to avoid needing.
This is why practitioners treat a predicted match as a hypothesis to test by synthesis
rather than as a result. It is also why we do not count spectral prediction among the routes
that could close the verification gap. It is still worth building, for a different reason. A
ranking that puts the correct structure in the top ten 92\% of the time on the NIST'20
[M+H]$^+$ subset \citep{wang2025iceberg2} is a good way to decide which candidate to spend a
synthesis on. That makes prediction an input to the allocation problem of
Section~\ref{sec:implications}, not an escape from it.

\subsection{Diffraction from tiny crystals}
\label{sec:microed}

Microcrystal electron diffraction is a genuine alternative, because it returns a structure
rather than a ranking. It solves structures from nanogram quantities of material that looks
amorphous, without deliberate crystal growth and without synthesis
\citep{jones2018microed}. Throughput is improving. Depositing HPLC fractions from crude
extracts as picolitre droplets puts 1{,}200 or more samples through a single autoloader
session \citep{delgadillo2024arrayed}. The approach has now given absolute stereochemistry
for newly discovered natural products, with cost and throughput named as what had held it
back \citep{delgadillo2025fungal}.

The bound is in the first word. The compound must form a crystal, at trace abundance, from a
complex matrix, at adequate purity. Polar, flexible, highly conjugated metabolites, which
describes much of what gut bacteria produce, are the least likely to oblige. So this route
exchanges one question for another. The issue is no longer whether the compound can be
synthesized, but whether it can be crystallized out of a mixture in which it is a minor
component. That is a
different wall in a more favourable position, not the absence of one.

\section{Sizing the gap}
\label{sec:measurable}

The argument so far is qualitative. It can be made quantitative, because both surrogate
routes have a measurable precondition. Synthesis needs a route, and diffraction needs a
crystal. We took the 193{,}611 HMDB entries flagged predicted or expected and
filtered them to small non-lipid structures, so as not to end up measuring combinatorial
lipid enumeration. We then sorted the remaining 10{,}645 compounds into four tiers by what
verifying each one would take (Figure~\ref{fig:tiers}; methods in
Appendix~\ref{app:methods}).

\begin{figure}[!tb]
  \centering
  \includegraphics[width=\linewidth]{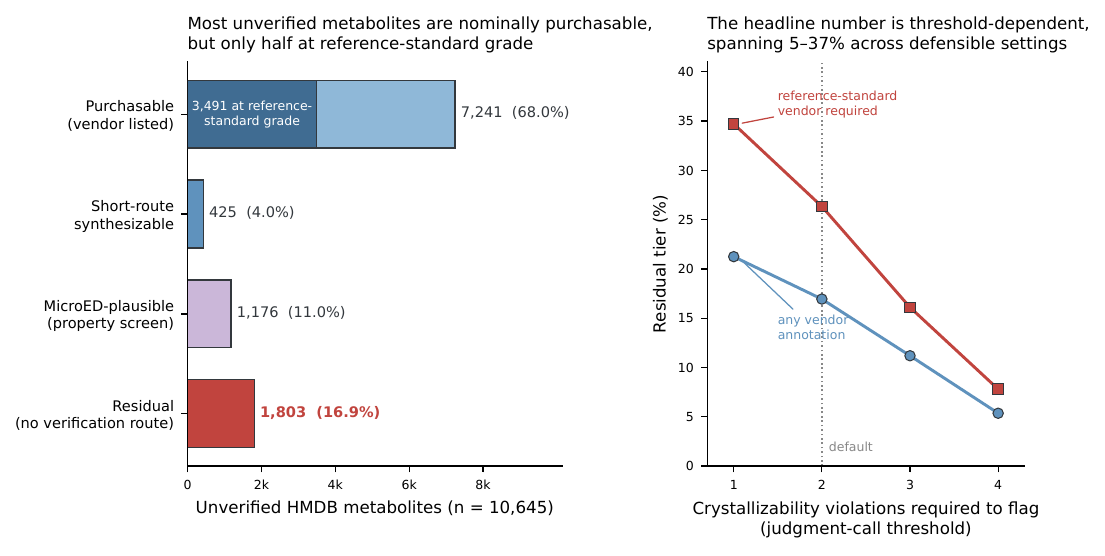}
  \caption{\textbf{Verification tiers over unverified HMDB metabolites.}
  The $n=10{,}645$ small non-lipid structures from the HMDB~5.0 \texttt{predicted} and
  \texttt{expected} classes, sorted by what confirming each would require
  (Appendix~\ref{app:methods}).
  \textbf{(left)} Tiers at default thresholds. Tier~1 counts any PubChem vendor
  annotation, which does not establish that a certified analytical standard exists. The dark
  segment inside that bar is the subset with a supplier of analytical-grade material, 3{,}491
  of the 7{,}241.
  \textbf{(right)} The residual spans 5.4--36.7\% across 72 threshold combinations. Within
  this set it is an upper bound, since 97.0\% of residual compounds are there because the
  retrosynthesis search found no route rather than because none exists; over the full
  candidate space it is a floor (Section~\ref{sec:measurable}).}\label{fig:tiers}
\end{figure}

Three caveats about the measurement belong here rather than in the appendix. The set is a
chemical filter over a human-wide database, not a curated gut metabolome. Tier~3 rests on a
property screen that has never been tested against crystallization outcomes, so it indicates
plausibility and not feasibility. And a compound recorded as having no synthetic route means
only that the planner found none. The same search failed on trimethylamine, which is
manufactured at megatonne scale. All three inflate the residual, and the sensitivity sweep in
the appendix reports the range, 5.4\% to 36.7\%, rather than a single figure.

A fourth caveat runs the other way, and it matters more. HMDB is a curated catalogue, so it
is a lower bound on the structures gut microbes actually make. A resource built specifically
for microbial metabolites already lists 29{,}295 of them across 3{,}725 species and strains
\citep{kruger2026mimedb2}, and both figures grow with each release. Generators are not
bounded by curation at all. Enumeration over known biotransformations
\citep{wishart2022biotransformer3} and iterative network expansion \citep{jeffryes2015mines}
produce candidates without limit. Our 10{,}645 compounds are therefore a small and
non-random sample of the real candidate space, drawn from the part of it that someone has
already thought worth cataloguing.

That biases our estimate downward. Section~\ref{sec:groundtruth} showed that availability
tracks prior attention, and the tier distribution shows the residual concentrating in the
entries a database itself marks as more speculative. Compounds absent from curated
databases are the extreme case of both. Nobody has studied them, so nobody sells them, and no
published route exists. We have no measurement for that population and do not claim one. But
the two gradients we can measure both point the same way, so the residual fraction we report
should be read as a floor on the gap over the full candidate space, not an estimate of it.

What the tiers show is less interesting than how they distribute. Split the set again by HMDB's status field. Of the better-established compounds, 16.0\% fall in the residual,
against 70.9\% of the more speculative ones ($n=182$). The difference holds on the
retrosynthesis axis alone, so it does not depend on the crystallizability thresholds. We
are cautious about the mechanism. The speculative compounds are also larger and more
flexible, which the property screen penalizes directly. Their structures are more unusual
too, which plausibly means a template library built from medicinal chemistry covers them
less well. Both effects are properties of our tools rather than of the
molecules.

That distinction matters less than it might appear. Either obtainability is genuinely lower for novel
structures, or our methods handle them worse. Operationally it comes to the same thing.
Verification is least available for exactly the compounds a discovery pipeline most wants
confirmed. The gap is not uniform across chemical space. It
is concentrated at the frontier.

\section{What follows}
\label{sec:implications}

We have argued that verification, not generation, is the binding constraint in this domain,
and that the gap is worst where discovery is most interesting. Two questions follow, and we
pose them without answering them.

The first is allocation. If only a fraction of predicted structures can ever be confirmed,
which ones should consume the synthesis and diffraction budget? Expected-information
arguments would favour compounds whose confirmation most reduces uncertainty about a
pathway. Cost arguments would favour whatever is cheapest to obtain, which as
Section~\ref{sec:measurable} shows means the compounds already closest to being known.
These criteria disagree, and the choice between them determines what gets confirmed. A
field that optimizes for the second while believing it optimizes for the first will
accumulate confirmations that add little information. We know of no principled treatment of this
tradeoff for structure elucidation. Active learning and Bayesian experimental design are
the usual answers to a question of this shape. They assume experiments are cheap, that many
rounds are possible, and that each result is informative about the cases left
untested. Confirmation offers none of the three. The allocation question is therefore real, but
it is not yet a form of the problem those methods solve.

The second is sufficiency. Level 1 is the standard, but almost nothing reaches it, so in
practice biologists already act on level 2 evidence. That is not obviously wrong. A
follow-up experiment that would be informative under either of two candidate structures
does not need the ambiguity resolved first, whereas a claim that a specific molecule
activates a specific receptor does. What is missing is any account of which downstream
decisions tolerate which level of structural confidence. Absent that account, the choice is
made implicitly, by whichever evidence happened to be available.

Automation is the most plausible route to narrowing this gap, and we expect it to work.
Retrosynthetic planning has been coupled to a robotically reconfigured continuous-flow
platform that runs machine-generated recipes \citep{coley2019robotic}. Language-model agents now plan and
execute syntheses end to end \citep{boiko2023coscientist, mbran2024chemcrow}, and a
substantial self-driving-laboratory literature has grown up around the problem
\citep{tom2024sdl}. Commercial parallel-synthesis hardware
routinely runs campaigns at a scale no manual laboratory would attempt
\citep{clarke2024htecrosscoupling}. A large enough reduction in cost per compound would
change the practical situation substantially, and nothing in our argument forecloses that.

What automation does not change is the unit of work. A reference standard confirms one
structure, because the two orthogonal measurements level~1 requires are made on one physical
sample. So verification cost stays proportional to the number of structures verified, while
generation cost does not. Faster synthesis lowers the constant; it does not
make confirmation amortizable the way fitting a model is. Purification, characterization and
the orthogonal measurements themselves also sit outside what current platforms automate. And
on-demand libraries are assembled from building blocks a supplier already stocks
\citep{tyrin2026onepotcore}, which limits what they can contain at any throughput.
The consequence is that the ordering problem does not dissolve with throughput. At any
capacity, some candidates get verified and others wait, so the question of which ones is
sharpened by better robots rather than retired by them.

Both questions generalize past metabolomics. Wherever candidate generation outpaces
confirmation, and confirmation is physical rather than computational, the same shape recurs.
There is a surrogate verifier of unquantified reliability, a scarce ground-truth verifier,
and no theory for allocating the second or calibrating trust in the first. Metabolomics is a clean test case. Its ground truth is
unambiguous when you have it, and you almost never have it.

\bibliographystyle{unsrtnat}
\bibliography{references}

\clearpage
\appendix
%
%

\section{Verification-tier analysis}
\label{app:methods}

\subsection{Data and subset selection}

Compounds were drawn from HMDB 5.0 \citep{wishart2022hmdb5}, parsed from the complete
\texttt{hmdb\_metabolites.xml} export (file timestamp 2021-10-23), giving 217{,}920
records and matching the entry count in the version paper. Version~5.0 remains the
current HMDB release, so that timestamp is the release date of the current data rather
than an old download. HMDB's \texttt{status} field
is our operational definition of verification state. Entries marked \texttt{expected}
(98{,}256) and \texttt{predicted} (95{,}355) are inferred from biochemistry or enzymology,
against \texttt{detected} (20{,}924) and \texttt{quantified} (3{,}385) entries resting on
experimental observation. The first two classes, 193{,}611 entries or 88.84\%, are the
analysis population.

That population cannot be used as-is. 87.4\% of it is combinatorial acyl-chain
enumeration, including 38{,}758 glycerophospholipids and 38{,}427 glycerolipids, and its
median relative molecular mass is 992 with a median of 52 rotatable bonds. A tier
analysis over it would measure lipid enumeration rather than the verification problem,
assigning nearly everything to the residual on flexibility alone. We therefore kept
unverified entries with $M_\mathrm{r} \le 500$ (RDKit, from deposited SMILES) that are
not ClassyFire \emph{Lipids and lipid-like molecules} and do not carry lipid shorthand
nomenclature, leaving \textbf{10{,}942 compounds} (10{,}760 \texttt{expected}, 182
\texttt{predicted}), with median $M_\mathrm{r}$ 268, 4 rotatable bonds, and $\log P$ 1.9. The
retained chemistry is dominated by organooxygen compounds, carboxylic acids, substituted
benzenes, flavonoids, isoflavonoids, coumarins, indoles and cinnamic acids, and several
thousand entries carry \emph{Food}, \emph{Plant} or \emph{Enteral} ontology annotations.

\paragraph{The gut filter is chemical, not annotation-based.} Gut-specific annotation is
almost absent from the unverified classes (64 \emph{Intestine}, 38 \emph{Microbe}), while
6{,}791 \texttt{detected}/\texttt{quantified} entries carry a \emph{Feces} annotation.
Provenance annotations derive from observation, so unobserved compounds have nothing to
annotate. Selecting on chemistry and dietary origin is the available proxy, and we report
it as a filter rather than a curated gut-metabolite set.

Deposited SMILES were standardized with RDKit \texttt{rdMolStandardize} (cleanup,
largest-fragment selection, charge neutralization, canonical tautomer) and keyed by
InChIKey. Three inputs failed standardization and 294 collapsed as duplicates, leaving
\textbf{10{,}645} unique structures; failures are typed and counted rather than dropped.

\subsection{Tier definitions}

Each compound is assigned to exactly one tier by the first matching rule, in the order
purchasable, a
supplier annotation exists (\S\ref{app:purch}); short-route synthesizable, a route of at
most \texttt{retro\_max\_steps\_for\_short\_route} steps to in-stock building blocks was
found or the compound is itself in stock (\S\ref{app:retro}); MicroED-plausible, not
flagged by the crystallizability screen (\S\ref{app:cryst}); residual, none of the above.
Every threshold is a named parameter in one configuration dictionary, and each compound
carries the rule that fired and the evidence for it, so assignments are auditable from
the released table alone.

\subsection{Purchasability}
\label{app:purch}

Standardized structures were resolved to PubChem CIDs (InChIKey, falling back to
canonical SMILES) and the PUG-View categorized-sources endpoint queried for suppliers
under \emph{Chemical Vendors}, rate-limited to 4~s$^{-1}$ with all responses cached.

\paragraph{A vendor annotation is not a certified standard.} Most listings are
research-grade material with no certificate of analysis, stated purity, or quantitative
reference value, whereas MSI level~1 needs a certified standard from a much smaller set
of suppliers. The annotation carries no stock, price or lead time, and a CID match may
correspond to a different stereoisomer, which matters for bile acids and other
metabolites whose epimers are distinct compounds. We therefore report a permissive
definition (any vendor) and a strict one (a curated list of analytical-standard houses).
The permissive definition is the default because it inflates tier~1 and so shrinks the
residual, the conservative direction for an argument that a gap exists. Even the strict
definition is a supplier-name heuristic, not a catalogue query.

\subsection{Retrosynthetic reachability}
\label{app:retro}

Routes were planned with AiZynthFinder~4.4.1 \citep{genheden2020aizynthfinder} run
locally, using the public USPTO expansion, ringbreaker and filter policies and the ZINC
stock file (17{,}422{,}831 InChIKeys) obtained via its \texttt{download\_public\_data}
mechanism, with a 120~s and 100-iteration budget per compound. ``In stock'' means
membership in that ZINC snapshot and nothing more, with no live vendor check, price or lead
time.

\paragraph{Search failure is a property of the tool, not the molecule.} Compounds with no
route found are recorded under a distinct status
(\texttt{NO\_ROUTE\_FOUND\_TOOL\_LIMITED}) and never labelled hard to synthesize. The
distinction is not cosmetic. During validation the planner found no route to
trimethylamine, manufactured at megatonne scale from methanol and ammonia, and failed on
essentially every bile acid tested including their taurine and glycine conjugates. Those
molecules are not synthetically difficult; their routes lie in C1 amination and steroid
semi-synthesis, which are under-represented in a template library derived from USPTO
medicinal chemistry. \textbf{The residual tier is therefore an upper bound on the true
verification gap}, and the tool-limited count is reported as a labelled category rather
than folded into it.

\subsection{Crystallizability screen}
\label{app:cryst}

Tier~3 applies a \emph{negative} property screen rather than predicting
crystallizability. RDKit descriptors ($M_\mathrm{r}$, cLogP, rotatable-bond count and its
size-normalized fraction, TPSA, H-bond donors and acceptors, aromatic rings, fraction
sp$^3$, Kier $\Phi$, estimated charge at physiological pH) were computed, and a compound
was flagged when it violated at least \texttt{cryst\_n\_violations\_for\_hard\_flag} of
the configured bounds.

\paragraph{The thresholds are transparent, not validated.} They were chosen to be legible
and sweepable, not because they predict crystallization outcome, and the screen has not
been tested against any crystallization dataset. Several bounds are judgment calls and
one is arguably signed wrongly. Strong directional hydrogen bonding, penalized here
through a donor-count bound, often \emph{promotes} crystal packing. The TPSA bound is
borrowed from oral-bioavailability heuristics about membrane permeation.
\citet{tang2021flexibility} further showed for para-substituted benzoic acids that
flexibility alone does not limit crystallizability, since a more flexible compound
nucleated faster than stiffer analogues, and that what matters is the conformational
energy barrier rather than the bond count. Better descriptors exist, including
\texttt{nConf20} \citep{wicker2016nconf20}; we compute and release Kier $\Phi$ but do not
threshold on it. Finally, MicroED solves structures from crystals orders of magnitude
smaller than single-crystal X-ray diffraction requires
\citep{jones2018microed, nannenga2026microed}, so tier~3 is a claim about physical
possibility, bounded in practice by instrument access rather than crystal size.

\subsection{Sensitivity analysis}

Because the residual depends on judgment-call thresholds, we report it as a range. We
swept 72 combinations of \texttt{cryst\_n\_violations\_for\_hard\_flag} $\in\{1,2,3,4\}$,
the purchasability definition, the minimum vendor count $\in\{1,2,5\}$, and the
short-route cutoff $\in\{2,3,5\}$. The residual spans \textbf{5.4\%--36.7\%} against
16.9\% at defaults (Figure~\ref{fig:tiers}, right). The table below gives marginal effects, with other parameters at their defaults.

\begin{center}
\begin{tabular}{llrrrr}
\toprule
parameter & value & tier 1 & tier 2 & tier 3 & residual \\
\midrule
\multirow{4}{*}{cryst.\ violations}
 & 1            & 7{,}241 &   425 & 718 & 21.2\% \\
 & 2 (default)  & 7{,}241 &   425 & 1{,}176 & 16.9\% \\
 & 3            & 7{,}241 &   425 & 1{,}788 & 11.2\% \\
 & 4            & 7{,}241 &   425 & 2{,}408 &  5.4\% \\
\midrule
\multirow{2}{*}{purchasability}
 & any vendor (default) & 7{,}241 &   425 & 1{,}176 & 16.9\% \\
 & reference standard   & 3{,}491 & 1{,}945 & 2{,}406 & 26.3\% \\
\midrule
\multirow{3}{*}{min.\ vendors}
 & 1 (default) & 7{,}241 &   425 & 1{,}176 & 16.9\% \\
 & 2           & 6{,}408 &   631 & 1{,}541 & 19.4\% \\
 & 5           & 5{,}291 & 1{,}008 & 1{,}929 & 22.7\% \\
\midrule
\multirow{3}{*}{max.\ route steps}
 & 2           & 7{,}241 &   340 & 1{,}212 & 17.4\% \\
 & 3 (default) & 7{,}241 &   425 & 1{,}176 & 16.9\% \\
 & 5           & 7{,}241 &   509 & 1{,}134 & 16.5\% \\
\bottomrule
\end{tabular}
\end{center}

Two features matter. The crystallizability threshold, the least defensible parameter, is
the most consequential, moving the residual by a factor of four. The route-step cutoff
barely matters (17.4\% to 16.5\%), because route length is not what places compounds in
the residual. For 97.0\% of them (1{,}749 of 1{,}803), no route was found at all. Tightening purchasability moves compounds out of tier~1 mostly into tiers~2 and~3.

\subsection{Throughput and reproducibility}

Retrosynthesis over all 10{,}645 compounds took 50~minutes on 160 workers with no engine
errors; 61 (0.6\%) hit the search cap. No sampling was needed, so the analysis is a census of
the filtered subset. PubChem lookups took 89~minutes; 68 structures (0.6\%) did not resolve to
a CID and 819 (7.7\%) returned no categorized-source record, both reported as distinct
categories rather than folded into the residual. Analysis code, the configuration dictionary, the per-compound tier table, and the
sensitivity sweep are released at
\url{https://github.com/tkoyama-keio/verification-wall-metabolomics}.

\end{document}